\documentclass[aps,pra,floatfix,superscriptaddress,twocolumn,showpacs,10pt]{revtex4-1}
\usepackage{amssymb, amsmath,color,mciteplus,graphicx,subfigure}
\usepackage{calc,epsfig,epstopdf,color,mciteplus,bm,mathrsfs}
\usepackage{times}

\begin{document}

\title{Realization of universal nonadiabatic holonomic quantum gates in trapped ion system}

\author{Yong-Sen Chen}
\affiliation{Key Laboratory of Atomic and Subatomic Structure and Quantum Control (Ministry of Education), and School of Physics, South China Normal University, Guangzhou 510006, China}

\author{Jing Gao}
\affiliation{Key Laboratory of Atomic and Subatomic Structure and Quantum Control (Ministry of Education), and School of Physics, South China Normal University, Guangzhou 510006, China}

\author{Li-Na Ji} \email{jlinamail@163.com}
\affiliation{Key Laboratory of Atomic and Subatomic Structure and Quantum Control (Ministry of Education), and School of Physics, South China Normal University, Guangzhou 510006, China}
\affiliation{Guangdong Provincial Key Laboratory of Quantum Engineering and Quantum Materials, \\ Guangdong-Hong Kong Joint Laboratory of Quantum Matter, and Frontier Research Institute for Physics, \\  South China Normal University, Guangzhou 510006, China}

\date{\today}

\begin{abstract}
The implementation of holonomic quantum computation is meaningful. We can effectively resist local and collective noise in the process of physical implementation by using the advantage of non-Abelian geometric phase. In this paper, we set out from the simplest and most fundamental Jaynes-Cummings model of ion trap system to implement single-qubit logical operations, and taking advantage of the interaction of two ions with a pair of laser pulses to implement two-qubit logical operations, respectively. Compared with the previous proposal, the whole process of our proposal are in tunable way and the universal holonomic quantum qubit gates can be made robust to systematic error and decay which pushes the gate fidelities in the presence of decoherence and systematic error to well high level. Thus our scheme affords an experimentally feasible and simple way to make realizing the robust nonadiabatic holonomic quantum computation.
\end{abstract}

\maketitle

\section{introduction}

Quantum computer has proved to be superior to the classical computer, but due to the fact that quantum states are susceptible to external interference, the practical conditions for the implementation of quantum computing are harsh. However, the recent development of holonomic quantum computation (HQC) has been a great solution to this problem \cite{FA,HQC,JP}. Because the geometric phase only depends on the overall geometry of the system evolution path, it can effectively resist local noise and can become a good medium for quantum computing \cite{berry,exp1,exp2,exp3,exp4,exp5,exp6,zhunoise,noise}. Duan et al.\cite{h3}, based on the ion trap system achieves adiabatic geometric quantum computation scheme by driving the quantum system to undergo appropriate adiabatic cyclic evolution. Since this scheme has to satisfy the condition of adiabatic theorem in the evolutionary process, it leads to the slow time evolution of the system, and the decoherence of the environment can be very large in actual operation. For this reason, the non-adiabatic geometric quantum computation scheme is more significant \cite{ZSL,ZSL1,XGF,ED1}.

The implementation of universal sets of robust high-speed geometric quantum gates by utilizing optical transitions in a generic three-level $\Lambda$ configuration in non-adiabatic evolution been proposed in Refs. \cite{XGF,ED1, vam2014, Zhang2014d, Xu2014, Xu2015, Xue2015b, Xue2016, Zhao2016, Herterich2016,xu2017, vam2017, Xue2017,xu20172,zhao2017,vam2016, 20,21,23}. By existing the cyclic evolution subspace in general quantum system can achieve non-adiabatic holonomic quantum computation (NHQC), and its scheme has been experimentally proved to be able to quickly implement the logic quantum gate and reducing the time evolution of the system on superconducting circuits \cite{Abdumalikov2013,Xu2018}, NMR \cite{Feng2013,li2017}, and  electron spins in diamond \cite{Zu2014, Arroyo-Camejo2014,nv2017}. In Refs. \cite{1,2,3}, developing a setting for conditional holonomic gates in a four-level configuration, and realizing non-adiabatic holonomic gates by coupled system consisting of quantum dots and single-molecule magnets, which offer promising scalability and robust efficiency.

In this paper, we set out from the simplest and most fundamental Jaynes-Cummings model of ion trap system and apply the aboved-mentioned four-level configuration scheme to implement single-qubit logical operations, and making use of the interaction of two ions with a pair of laser pulses to implement two-qubit logical operations, respectively. Compared with the previous proposal, the whole process of our proposal are in tunable way and the universal holonomic quantum qubit gates can be made robust to systematic error and decay which pushes the gate fidelities in the presence of decoherence and systematic error to well high level. Thus our scheme affords an experimentally feasible and simple way to make realizing the robust nonadiabatic holonomic quantum computation.

The outline of the paper is organized as follows. In section II, we demonstrate a universal set of non-adiabatic single-qubit holonomic gates by taking advantage of the simplest and most fundamental Jaynes-Cummings model of ion trap system. In section III, we will illustrate the construction of two-qubit logical gate, and then taking holonomic phase gate and a two-qubit controlled-phase gate as examples to show that these gates can be made robust to decay and systematic error in section IV. The paper ends with the conclusions.

\section{Universal single-qubit gates}
\begin{figure}
  \centering
  \includegraphics[width=8.5cm]{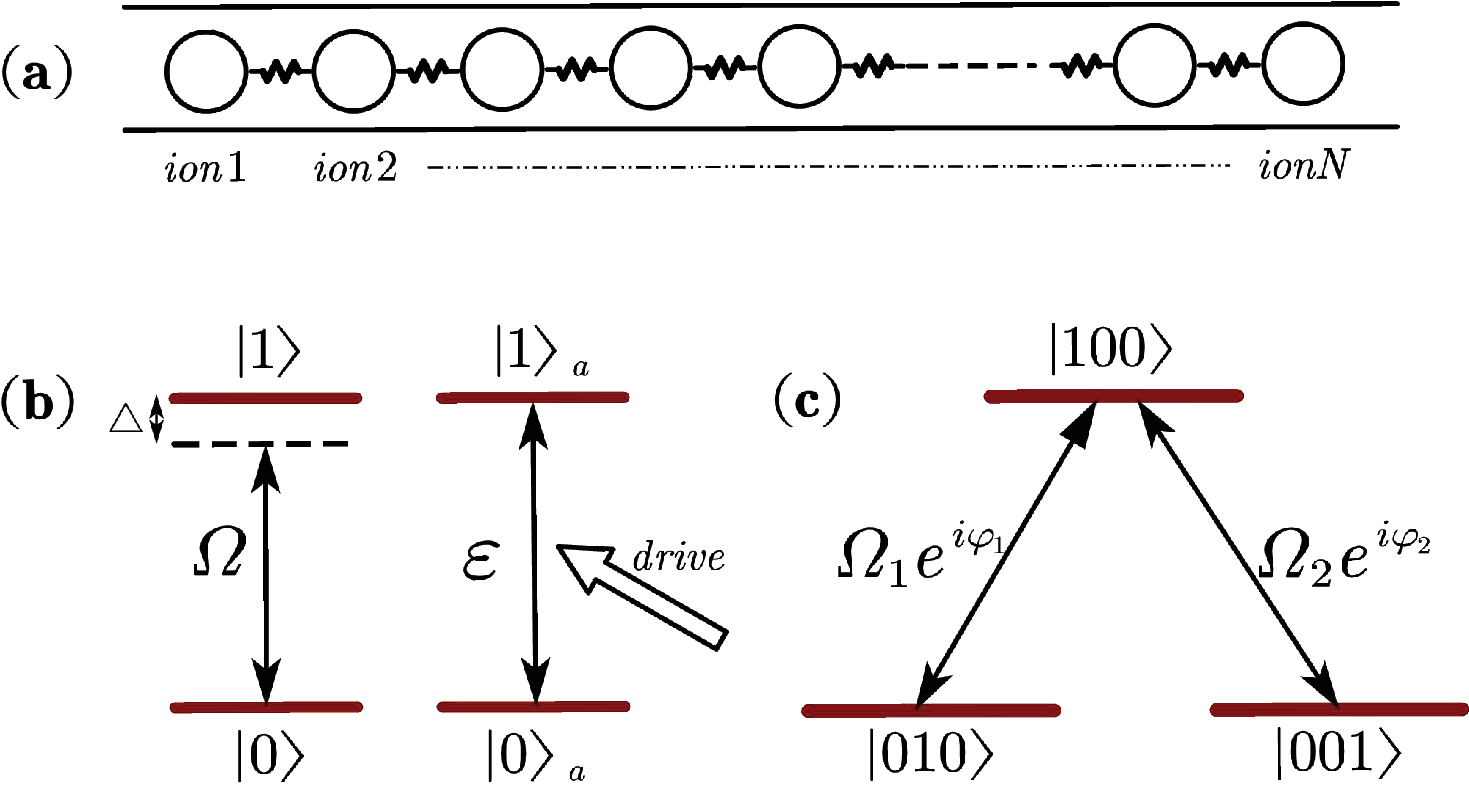}\\
  \caption{(a) Considering a string of N two-level ions in trap, which can be individually addressed by the laser pulse. (b) The interaction of a two-level ion with the laser field in the ion trap, and adding the external driving of the collective vibrational quantum bit. (c) Equivalent energy level and driving configuration for the two qubit gate. Applying the interaction of two ions with a pair of laser pulses to implement two-qubit logical operations.}\label{Fig single11}
\end{figure}

Setting out from the simplest and most fundamental Jaynes-Cummings model of ion trap system to establish the universal set of nonadiabatic holonomic single-qubit quantum gates. We consider the interaction of a two-level ion with a laser field, and adding the external driving of the collective vibrational quantum bit, as shown in Fig.\ref{Fig single11}(b). The associated Hamiltonian reads
\begin{eqnarray}
\label{Eq1}
H_1&=&\omega_\nu a^\dagger a+\frac{1}{2}\omega\sigma_z+\Omega_0e^{i\eta(a^\dagger+a)}\sigma^\dagger e^{-i(\omega_\text{L} t-\phi^\prime)}\notag \\
&&+\frac{\varepsilon}{2} a^\dagger e^{-i\omega_\text{d}t}+\text{H.c.}
\end{eqnarray}
where $\eta$ is the Lamb-Dicke parameter, $\sigma_z=|1\rangle\langle1|-|0\rangle\langle0|$, $\sigma=|0\rangle\langle1|$ and $\sigma^\dagger=|1\rangle\langle0|$ are the standard Pauli operators down and up, respectively, associated with the two-level system with transition frequency $\omega$ of an ion; $a$ and $a^\dagger$ are the annihilation and creation operators for the collective vibrational mode, $\omega_\nu$ is the frequency of vibrational mode; $\phi^\prime$ and $\omega_L$ denotes the phase and frequency of the laser field, respectively; $\omega_d$ is the frequency of the external driving of the collective vibrational quantum bit; $\Omega_0$ and $\varepsilon$ are the coupling constants of the interaction of ion with the the laser pulse and the collective vibrational quantum bit with the driving field, respectively.

We apply the rotating-wave approximation by neglecting the oscillating terms in the interaction picture, and then making $\Delta=\omega-\omega_L=\omega_\nu, \omega_\nu=\omega_d$. The interaction Hamiltonian under the consideration of Lamb-Dicke limit takes its final form as
\begin{eqnarray}
\label{Eq2}
H_1&=&\Omega_0^\prime a^\dagger \sigma e^{-i\phi}+\frac{\varepsilon}{2} a^\dagger+\text{H.c.}
\end{eqnarray}
where $\Omega_0^\prime=\eta\Omega_0,\phi=\phi^\prime+\pi/2$. Due to the different motional states of the collective vibrational mode can realize large nonlinear, so that it can separately control vacuum motional state and the first motional number state to encode a collective vibrational quantum bit, and will hardly inspire other motional state energy level transition \cite{4}. So the Hamiltonian of the system can be expanded by a four-dimensional effective state space
\begin{eqnarray}
H_1=
\left(
\begin{array}{cccc}
 0 & 0 & \frac{\varepsilon}{2} & 0 \\
 0 & 0 & \Omega_0^\prime e^{i\phi}  & \frac{\varepsilon}{2} \\
\frac{\varepsilon}{2} & \Omega_0^\prime e^{-i\phi}  & 0 & 0 \\
 0 & \frac{\varepsilon}{2} & 0 & 0 \\
\end{array}
\right)
\end{eqnarray}
in the ordered orthonormal basis $M=\{|00\rangle,|01\rangle,|10\rangle,|11\rangle\}$
where $|mn\rangle\equiv |m\rangle_\text{a}\otimes|n\rangle_\text{q}$, i.e., they denote the external motional state (be used as auxiliary control qubit) and the internal electronic state (be used as target qubit) of ion, respectively. In a nutshell, we will implement the construction of a series of holonomic single-qubit gates in a four-level structure. By resetting the parameters $\varepsilon=J\sin\frac{\theta}{2}$, $\Omega_0^\prime=J\cos\frac{\theta}{2}$, the Hamiltonian can be reduced to
\begin{eqnarray} \label{Eq4}
H_1=
{J} \left(
\begin{array}{cc}
0 & T \\
T^\dagger & 0 \\
\end{array}
\right)
\end{eqnarray}
where
\begin{eqnarray}
T=
\left(
\begin{array}{cc}
\frac{\sin(\theta/2)}{2} & 0 \\
\cos\frac{\theta}{2} e^{i\phi}  & \frac{\sin(\theta/2)}{2} \\
\end{array}
\right)
\end{eqnarray}
Here we find matrix $T$ is a complex-valued and time-independent $2\times2$ matrix so that it can satisfies condition $\text{det}T\neq0$ and ensure that matrix $T$ is invertible, thus there is a unique singular value decomposition of matrix $T=WD V^\dagger$, where

\begin{eqnarray}
W=
\left(
\begin{array}{cc}
\sin\frac{\theta}{4} & \cos\frac{\theta}{4}e^{-i\phi} \\
 \cos\frac{\theta}{4}e^{i\phi} & -\sin\frac{\theta}{4} \notag\\
\end{array}
\right),
D=
\left(
\begin{array}{cc}
\cos^2\frac{\theta}{4} & 0 \\
0 & \sin^2\frac{\theta}{4} \\
\end{array}
\right)
\end{eqnarray}
\begin{eqnarray}
V=
\left(
\begin{array}{cc}
\cos\frac{\theta}{4} & \sin\frac{\theta}{4}e^{-i\phi} \\
\sin\frac{\theta}{4}e^{i\phi} & -\cos\frac{\theta}{4} \\
\end{array}
\right)
\end{eqnarray}
respectively. By applying the above matrix singular value decomposition method and the four-dimensional effective state space of the Hamiltonian $H_1$ are separated into two two-dimensional subspace, i.e.,
\begin{eqnarray}
M=M_0\oplus M_1
\end{eqnarray}
where $M_0=\{|00\rangle,|01\rangle\}$, and $M_1=\{|10\rangle,|11\rangle\}$, this implies that in the basis $M$ the time evolution
operator splits into $2\times2$ blocks, the final time evolution operator can be expressed as
\begin{eqnarray}
U(t)=
\left(
\begin{array}{cc}
W\cos(a_\text{t}D)W^\dagger & -iW\sin(a_\text{t}D)V^\dagger \\
-iV\sin(a_\text{t}D)W^\dagger & V\cos(a_\text{t}D)V^\dagger \\
\end{array}
\right)
\end{eqnarray}
where $a_\text{t}=\int^\text{t}_0 J(t')dt'$.

We demonstrate in the following how two path segments are sufficient to realize arbitrary holonomic one-qubit gates. In the first path segment $[0,\tau_1]$, we define $T_1=W_1D_1V^\dagger_1$, and choosing parameters such that $\cos(a_{\tau_1} D_1)=0$, $\sin(a_{\tau_1} D_1)=P^1_i$ , where $i=I,z$, and $P^1_I=\text{diag}\{1,1\}$, $P^1_z=\text{diag}\{1,-1\}$ (In order to satisfy this equality constraint, it need to limit the parameter $\theta\neq n\pi$, where $n$ is an integer), we can obtain the first path segment's time evolution operator
\begin{eqnarray}
U(C_1)=
\left(
\begin{array}{cc}
0 & -iW_1P^1_i V^\dagger_1 \\
-iV_1P^1_iW^\dagger_1 & 0 \\
\end{array}
\right)
\end{eqnarray}
where the final selection of the value of $i$ depends on the change of the parameters $(\theta,\phi_1,a_{\tau_1})$. In the second path segment $[\tau_1,\tau_2]$, we take the same parameter $(\theta,a_{\tau_2})$, only to change the relative phase parameter $\phi_2$, so that the path evolves back along the second path segment. In the same way, we define $T_2=W_2D_2V^\dagger_2$, and choosing parameters such that $\cos(a_{\tau_2}D_2)=0$, $\sin(a_{\tau_2}D_2)=P^2_i$, where $i=I,z$, and $P^2_I=\text{diag}\{1,1\}$, $P^2_z=\text{diag}\{1,-1\}$, we obtain the second path segment's time evolution operator
\begin{eqnarray}
U(C_2)=
\left(
\begin{array}{cc}
0 & -iW_2P^2_i V^\dagger_2 \\
-iV_2P^2_iW^\dagger_2 & 0 \\
\end{array}
\right)
\end{eqnarray}
Through consecutive application of $U(C_1)$ and $U(C_2)$ can generates a loop $C_1\ast C_2$, we obtain the final time evolution operator
\begin{eqnarray}
U=U(C_1\ast C_2)
=|0\rangle_\text{a}\langle0|\otimes U_0+|1\rangle_\text{a}\langle1|\otimes U_1
\end{eqnarray}
where the evolution operator $U_0=-W_2P^2_i V^\dagger_2V_1P^1_iW^\dagger_1$ and $U_1=-V_2P^2_iW^\dagger_2W_1P^1_i V^\dagger_1$ act on the target qubit by conditionalized on the states $|0\rangle_\text{a}$ and $|1\rangle_\text{a}$ of the control qubit, respectively.

Clearly, we can find the whole process are nonadiabatic holonomy transformations: (i) by considering the evolution of the orthogonal subspaces $M_0$ and $M_1$ in each path segment, one may notice that the two unitaries $U_0$ and $U_1$ are purely geometric since the Hamiltonian $H$ vanishes on $M_0$ and $M_1$ separately; (ii) the evolution of the orthogonal subspaces $M_0$ and $M_1$ also undergoes cyclic evolution, respectively. For example, in the two-dimensional subspaces $M_0$ ($M_1$), satisfying the condition of cyclic evolution since $U_0(|00\rangle\langle00|+|01\rangle\langle01|)U^\dagger_0
=|00\rangle\langle00|+|01\rangle\langle01|$
 ($U_1(|10\rangle\langle10|+|11\rangle\langle11|)U^\dagger_1
=|10\rangle\langle10|+|11\rangle\langle11|$).

Based on the feasibility of the experiment, we can prepare the control qubit in vacuum motional state $|0\rangle_\text{a}$, i.e., selecting the evolution operator $U_0$ in orthogonal subspace $M_0=\{|00\rangle,|01\rangle\}$. Letting $P^1_i\equiv P^1_I=\text{diag}\{1,1\}$, $P^2_i\equiv P^2_I=\text{diag}\{1,1\}$ and choosing $\phi_2=\pi$, $\phi_1=0$ in two path segments, we can obtain the time evolution operator
\begin{eqnarray}
U^\prime_0=
\left(
\begin{array}{cc}
 \cos\theta & -\sin\theta \\
\sin\theta & \cos\theta \\
\end{array}
\right)
=e^{-i\theta \sigma_y}
\end{eqnarray}
So we get the rotation matrix around the Y-axis by an angle $\theta$; adjust the parameters, if we take $P^1_i\equiv P^1_z=\text{diag}\{1,-1\}$, $P^2_i\equiv P^2_z=\text{diag}\{1,-1\}$ in two path segments, similarly above, the time evolution operator is
\begin{eqnarray}
U^{\prime\prime}_0=e^{i\pi}
\left(
\begin{array}{cc}
 e^{-i \Delta\phi} & 0 \\
0 & e^{i \Delta\phi} \\
\end{array}
\right)=e^{i\pi} e^{-i\Delta\phi \sigma_z}
\end{eqnarray}
where $\Delta\phi=\phi_2-\phi_1$, ignoring the global phase factor $e^{i\pi}$, finally we can get the rotation matrix around the X-axis by an angle $\Delta\phi$, any phase gate can be achieved. To sum up, through the choice of the external vacuum motional state, we can apply internal electronic states $|0\rangle$ and $|1\rangle$ to construct a series of universal single-qubit gates.

\section{The two-qubit gate}

We next turn to achieve the nontrivial holonomic two-qubit gate by the use of the internal ions' qubits state. Combining the implemented arbitrary single-qubit gates, universal quantum computation can be realized. We utilize a pair of laser pulses with frequencies $\omega_1-\omega_\nu$ (phase $\varphi_1$), $\omega_2-\omega_\nu$ (phase $\varphi_2$), respectively. Here $\omega_0$($\omega_1$) is the resonant frequency between the two energy states $|1\rangle$ and $|0\rangle$ of the $1$($2$)th ions.
Considering the interaction of two ions with the above laser pulses, where two ions are in the same collective vibrational mode in ion trap. The same calculation process as the single-qubit process, the corresponding Hamiltonian is
\begin{eqnarray}
H_2=\Omega_1 \sigma_1^+ a e^{i\varphi_1} +\Omega_2\sigma_2^+ a e^{i\varphi_2} +\text{H.c.}
\end{eqnarray}
where $\sigma_j$ and $\sigma^\dagger_j$ $(j=1,2)$ are the Pauli operators down and up of the $j$th ion, respectively. $\Omega_j$ is the coupling constants of the interaction of the $j$th ion with the laser pulses. In the ordered orthonormal basis $B=\{|100\rangle,|010\rangle,|001\rangle\}$,
where $|\alpha\beta\gamma\rangle\equiv |\alpha\rangle_\text{a}\otimes|\beta\rangle_{q_1}\otimes|\gamma\rangle_{q_1}$, i.e., they denote the same external motional state (be used as auxiliary control qubit) and target qubits of the first ion and the second ion, respectively, the Hamiltonian $H_2$ can simply reduces to
\begin{eqnarray} \label{me}
H_2=
\left(
\begin{array}{cccc}
 0 & \Omega_1^\ast e^{-i\varphi_1}  & \Omega_2^\ast e^{-i\varphi_2}  \\
 \Omega_1 e^{i\varphi_1} & 0 & 0   \\
 \Omega_2 e^{i\varphi_2} & 0  & 0  \\
\end{array}
\right)
\end{eqnarray}
By resetting the parameters $\Omega_1=\Omega\cos\frac {\vartheta} {2}$ and $\Omega_2=\Omega\sin\frac {\vartheta} {2}$ where $\Omega=\sqrt{\Omega_1^2+\Omega_2^2}$. The above formula thus establishes in $B$ a $\Lambda$-type Hamiltonian, equivalent energy level and driving configuration as shown in Fig.\ref{Fig single11}(c). For simplicity, this can be illustrated in the dressed-state representation where the two lowest states of this three-level system are
\begin{eqnarray}
|d\rangle&=&\sin\frac{\vartheta}{2}|010\rangle-\cos\frac{\vartheta}{2}
e^{i\varphi}|001\rangle\\
|b\rangle&=&\cos\frac{\vartheta}{2}|010\rangle+\sin\frac{\vartheta}{2}e^{i\varphi}|001\rangle
\end{eqnarray}
where $\varphi=\varphi_2-\varphi_1$. Then the Eq.(\ref{me}) can be rewritten as
\begin{eqnarray}
H_2=\Omega(e^{i\varphi_1} |b\rangle \langle100|+e^{-i\varphi_1} |100\rangle \langle b|)
\end{eqnarray}
We demonstrate in the following how two path segments are sufficient to realize the construction of holonomic two-qubit gates by the use of the internal ions¡¯ qubits state. For the first segment, $\int^{\frac{\tau}{2}}_0\Omega(t) dt =\frac{\pi}{2}$ and $\varphi_1=0$; and the second segment, $\int^\tau_{\frac{\tau}{2}}\Omega(t) dt =\frac{\pi}{2}$ and $\varphi_1=\chi$. Here $\tau$ represents the runtime of the whole this operation. Obviously, the dark state $|d\rangle$ is decoupled from the other states, while the bright state $|b\rangle$ is coupled to the excited state $|100\rangle$ with effective Rabi frequency $\Omega(t)$. Because of $\int^\tau_0\Omega(t) =\pi$, the dressed states undergo a cyclic evolution in which $| d \rangle$ remains invariant and $| b \rangle$ evolves to $-| b \rangle$. Moreover, as $\langle\psi_{\text{i}}(t)|H_2|\psi_{\text{j}}(t)\rangle=0$ with $|\psi_{\text{i,j}}\rangle\in \{|001\rangle, |010\rangle\}$, there is no transition between the two time-dependent states, i.e., the evolution satisfies the parallel-transport condition. Therefore, the evolution operator can be represented as
\begin{eqnarray*}
U_2&=&e^{-i\int^\tau_{\frac{\tau}{2}}\Omega(t)(e^{i\chi} |b\rangle \langle100|+\text{H.c.})dt} e^{-i\int^{\frac{\tau}{2}}_0\Omega(t)(|b\rangle \langle100|+\text{H.c.})dt}  \notag\\
   &=&|d\rangle \langle d|-(e^{i\chi} |b\rangle \langle b|+e^{-i\chi} |100\rangle \langle 100|)
\end{eqnarray*}
which it can realize the holonomic operations under the above two conditions. Similarly, we prepare the control qubit in vacuum motional state $|0\rangle_\text{a}$. We can apply internal electronic states of two ions to realize arbitrary two-qubit controlled-phase gate by choosing the parameters $\vartheta=\pi$, finally such evolution can be represented in the four-dimensional subspace $\mathrm{span}\left\{ |00\rangle, |01\rangle, |10\rangle, |11\rangle\right\}$ (i.e., they denote the states of two ions $|\beta\gamma\rangle_{q_1 q_2}$\},
\begin{eqnarray}
U_2=
\left(
\begin{array}{cccc}
 1 & 0 & 0 & 0  \\
 0 & 1 & 0 & 0  \\
 0 & 0 & 1 & 0  \\
 0 & 0 & 0 & -e^{i\chi}  \\
\end{array}
\right)
\end{eqnarray}
and then we utilize the above elementary gate $U_2$ to realize composite gate $U_2U_2$  \cite{11,22} not only preserving holonomic robustness but also suppressing systematic errors when considering the effect of decoherence. In the next section, we will demonstrate it by the numerical simulation. Obviously, the form of arbitrary two-qubit controlled-phase gate $U_2U_2$ also can be implemented. Therefore, in general, a nontrivial holonomic two-qubit gate can be achieved.

\section{Decoherence and systemic error}
\begin{figure}
  \centering
  \includegraphics[width=8.5cm]{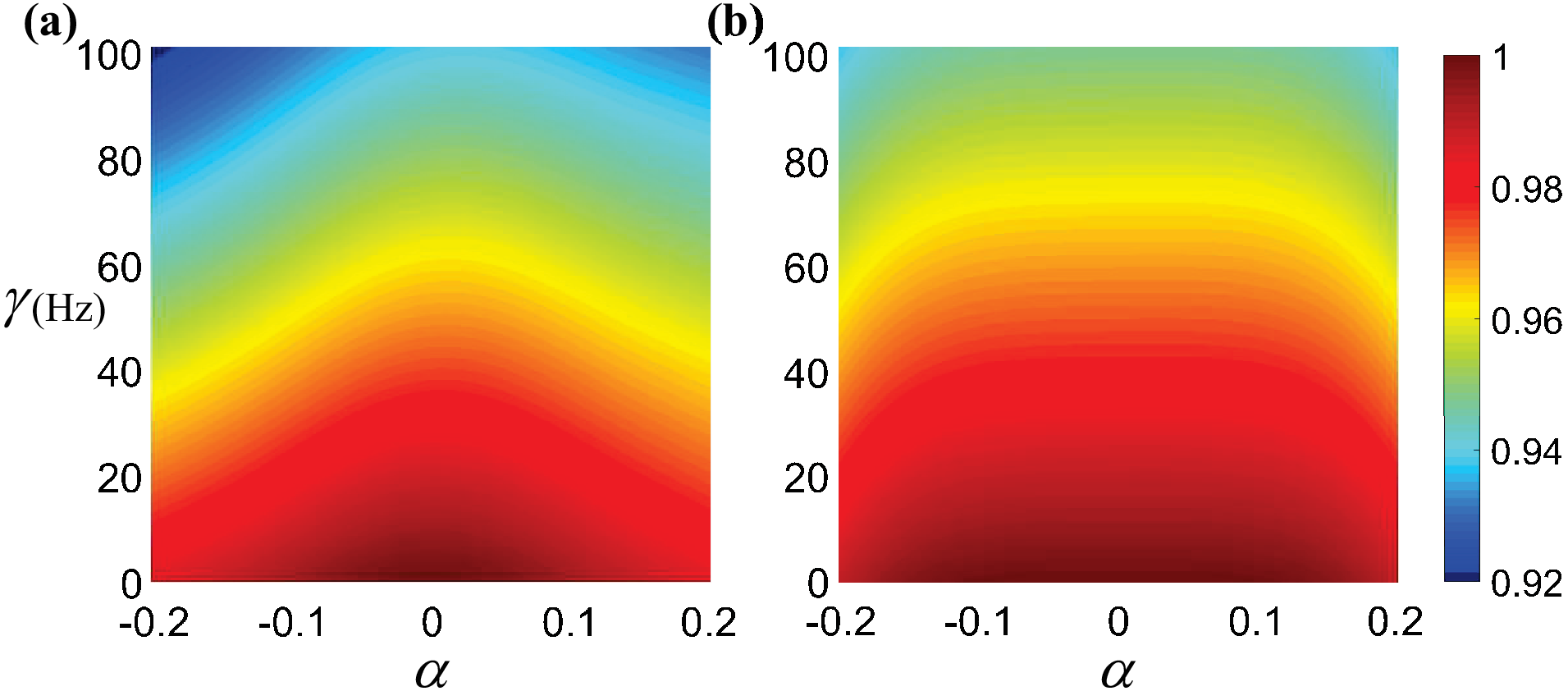}\\
  \caption{The fidelities $F$ versus the systematic error of effective coupling strength and decoherence strength. The range of the corresponding parameter values $\alpha$ and $\gamma$ are $-0.2\sim0.2$ and $0Hz\sim100Hz$, respectively. (a) The situation of holonomic single-qubit phase gate. (b) The situation of two-qubit controlled-phase gate. }\label{Fig single111}
\end{figure}

The imperfection of a quantum gate is usually due to the environment of physical implementation and the fluctuations of the control fields , we now turn to take holonomic single-qubit phase gate and a two-qubit controlled-phase gate as examples to show the performance of these gates when considering decoherence and systematic error.

We firstly need to know the form of systematic error which is inevitable in the course of the experiment. For the Hamiltonian $H_1$ and $H_2$, the most unstable controlling parameters are the coupling strength of the interaction of the ion with the laser pulse. Here we consider the error of the effective coupling strength, the corresponding form is $(1+\alpha)J$ and $(1+\alpha)\Omega$, respectively, where $\alpha$ represent the unknown time-independent fractions of laser fields.

The decoherence process is also unavoidable in the process of physical implementation, understanding its effect in our scheme becomes a crucial part. Considering the main decoherence effect, the evolution of the system in the Markov approximation is governed by the following Lindblad master equation:
\begin{eqnarray}
\label{master1}
\dot\rho&=& -i[H_k, \rho]+\frac \kappa 2\mathscr{L}(a)+\frac {\gamma _-} {2}( \mathscr{L}(\sigma^1)+ \mathscr{L}(\sigma^2))\notag \\
        && + \frac {\gamma _{z}} {2} (\mathscr{L}(\sigma^1 _z)+\mathscr{L}(\sigma^2 _z))
\end{eqnarray}
where $H_k$($k=1,2$) is the full Hamiltonian given above when considering systematic error , $\rho$ is the density matrix, $\mathscr{L}(\mathcal{A})=2\mathcal{A}\rho
\mathcal{A}^\dagger-\mathcal{A}^\dagger \mathcal{A} \rho -\rho \mathcal{A}^\dagger \mathcal{A}$ is the Lindbladian of the operator $\mathcal{A}$, and $\kappa$, $\gamma _-$, and $\gamma _z$ are the decay rate of the vibrational mode, and the relaxation and dephasing rates of the $j$th ion qubit, respectively. Here we choose holonomic single-qubit phase gate and a two-qubit controlled-phase gate as two typical examples, the fidelities of which are defined as $F=[tr(\rho_\text{final} \rho_\text{ideal})]^{1/2}$, where $\rho_\text{final}(\rho_\text{ideal})$ being the density operator of the system in the case with (without) decoherence process and systematic error.

We choose the experimentally achievable parameters as the following \cite{NP}: $\eta=0.044$, $|\varepsilon|\simeq2\pi\times2.42$ \text{kHz},  $|\Omega_0^\prime|\simeq2\pi\times4.20$ \text{kHz}, the holonomic single-qubit phase gate can be realized; $|\Omega_1|=0$ \text{kHz}, $|\Omega_2|\simeq2\pi\times4.84$ \text{kHz}, the two-qubit controlled-phase gate can be realized. For simplicity, we set $\kappa$, $\gamma_-$, $\gamma_z$ are all on the same order of magnitude, i.e. $\gamma=\kappa=\gamma_-=\gamma_z$. From figure \ref{Fig single111}, we can shows the simulated fidelities of the holonomic single-qubit phase gate and controlled-phase gate as a function of the systematic error (the range of parameter values $\alpha$: $-0.2\sim0.2$) and decoherence strength (the range of parameter values $\gamma$: $0\sim100$ \text{Hz}). The analysis here implies that these gates can be made robust to systematic error and decay.

\section{Conclusion}

In summary, we haved proposed to implement the NHQC in trapped ion system. We set out from the simplest and most fundamental Jaynes-Cummings model of ion trap system and the interaction of two ions with a pair of laser pulses to implement single-qubit and two-qubit logical operations, respectively. Compared with the previous proposal, the whole process of our proposal are in tunable way and the universal holonomic quantum qubit gates can be made robust to systematic error and decay which pushes the gate fidelities in the presence of decoherence and systematic error to well high level. Therefore, our scheme afford a feasible and simplified way to make realizing the robust NHQC in trapped ion system, which are promising candidates for quantum computation.

\bigskip

\acknowledgments

This work was supported by the Fund for Young Investigators from SCNU (Grant No. 2024KJ29), and National College Students' Innovation Training Program (Grant No. 202310574056).

\end{document}